\documentclass{article}
\usepackage[a4paper, total={6.5in, 10in}]{geometry}

\usepackage[utf8]{inputenc}
\usepackage{csquotes}
\usepackage[numbers, sort&compress]{natbib}
\usepackage{comment}
\usepackage{xcolor}
\usepackage{authblk}
\usepackage{hyperref}
\usepackage{graphicx}
\usepackage{booktabs}
\usepackage{verbatim}
\usepackage{tikz}
\usetikzlibrary{shapes.geometric, arrows.meta, fit,positioning,arrows, positioning,quotes, calc, backgrounds}

\tikzset{%
	>={Latex[width=2mm,length=2mm]},
	base/.style = {rectangle, rounded corners, draw=black,
		minimum width=4cm, minimum height=1cm,
		text centered, font=\sffamily},
	activityStarts/.style = {base, fill=blue!30},
	startstop/.style = {base, fill=gray!30},
	activityRuns/.style = {base, fill=gray!30},
	process/.style = {base, minimum width=2.5cm, fill=orange!15,
		font=\ttfamily},
}

\usepackage{soul}

\hypersetup{
    colorlinks=true,
    linkcolor=blue,
    filecolor=magenta, 
    citecolor=blue,
    urlcolor=cyan,
    pdftitle={AI to improve clinical coding practice in Scandinavia},
    pdfpagemode=FullScreen,
    }

\title{Artificial intelligence to improve clinical coding practice in Scandinavia: a crossover randomized controlled trial}

\author[1,2]{Taridzo Chomutare}
\author[1]{Therese Olsen Svenning}
\author[1,3]{Miguel Ángel Tejedor Hernández}
\author[1,4]{Phuong Dinh Ngo}
\author[1,4]{Andrius Budrionis} 
\author[1]{Kaisa Markljung}
\author[5]{Lill Irene Hind}
\author[1]{Torbjørn Torsvik}
\author[4,6]{Karl Øyvind Mikalsen}
\author[8]{Aleksandar Babic}
\author[1,7]{Hercules Dalianis}

\affil[1]{Health Data Analytics, Norwegian Centre for E-health Research, Tromsø, Norway}
\affil[2]{Department of Computer Science, UiT The Arctic University of Norway, Tromsø, Norway}
\affil[3]{Department of Mathematics and Statistics, UiT The Arctic University of Norway, Tromsø, Norway}
\affil[4]{Department of Physics and Technology, UiT The Arctic University of Norway, Tromsø, Norway}
\affil[5]{Clinic for Surgery, Oncology and Women Health (K3K), University Hospital of North Norway, Tromsø, Norway}
\affil[6]{The Norwegian Centre for Clinical Artificial Intelligence (SPKI), University Hospital of North Norway, Tromsø, Norway}
\affil[7]{Department of Computer and Systems Sciences, Stockholm University, Kista, Sweden}
\affil[8]{Group Research and Development, DNV, Høvik, Norway}

\date{July 2025}

\begin{document}
\maketitle

\begin{abstract}
\textbf{Trial design} Crossover randomized controlled trial.
\textbf{Methods} An AI tool, Easy-ICD, was developed to assist clinical coders and was tested for improving both accuracy and time in a user study in Norway and Sweden. Participants were randomly assigned to two groups, and crossed over between coding complex (longer) texts versus simple (shorter) texts, while using our tool versus not using our tool.
\textbf{Results} Based on Mann-Whitney U test, the median coding time difference for complex clinical text sequences was 123 seconds (\emph{P}\textless.001, 95\% CI: 81 to 164), representing a 46\% reduction in median coding time when our tool is used. There was no significant time difference for simpler text sequences. For coding accuracy, the improvement we noted for both complex and simple texts was not significant.  
\textbf{Conclusions} This study demonstrates the potential of AI to transform common tasks in clinical workflows, with ostensible  positive impacts on work efficiencies for complex clinical coding tasks. Further studies within hospital workflows are required before these presumed impacts can be more clearly understood.\\ 


\end{abstract}

\section{Introduction}

\subsection{Background}

International Statistical Classification of Diseases and Related Health Problems codes, tenth revision (ICD-10) \cite{Moriyama2011-in} play an important role in healthcare. All hospitals in Scandinavia record their activity by summarizing patient encounters into ICD-10 codes. Clinical coding directly affects how health institutions function on a daily basis because they are partially reimbursed based on the codes they report. The same codes are used to measure both volume and quality of care, thereby providing an important foundation of knowledge for decision makers at all levels in the healthcare service. 

Clinical coding is a highly complex and challenging task that requires a deep understanding of both the medical terminology and intricate clinical documentation. Coders must accurately translate detailed patient records into standardized codes, navigating the inherently complex medical language, which make this task prone to errors and inconsistencies. Continuously evolving, and progressively complex coding standards in healthcare, coupled with human factors, have been shown to lead to documentation of poor quality, which in turn often leads to inappropriate treatment and follow up of patients \cite{stanfill2014preparing,stausberg2008reliability}. 

Unlike in some parts of the world, like USA and UK, where clinical coding is taken as a serious profession with advanced training curriculum, in Scandinavia, the common practice is for junior doctors to perform the coding and for medical secretaries to quality-assure the coding. This operational workflow difference may be a partial explanation for why clinical coding practice in Scandinavian hospitals is unsatisfactory. Investigations from Norway \cite{Mathisen2015-jq} and from the other countries  
\cite{jacobsson_serden2013_socialstyrelsen,Schmidt2015-ir,Stegman2005-dk,So2010-yw} have exposed similar issues. The extent of these coding issues will vary across countries and medical areas. For instance, the Swedish National Board of Health and Welfare reports that, in hospital cancer care, 22\% of the main diagnosis were wrong \cite{jacobsson_serden2013_socialstyrelsen}, while in Norway, the Office of the Auditor General of Norway (Riksrevisjonen) reported that up to 40\% of the main diagnoses for lung diseases were incorrectly coded when reviewing selected patient records from 2017 \cite{riksrevisjonenUndersxF8kelseMedisinsk}. 

Computer-Assisted Coding (CAC) has been proposed a plausible tool for solving these problems. The American Health Information Management Association defines CAC as the “use of computer software that automatically generates a set of medical codes for review, validation, and based upon provider clinical documentation” \cite{ahima2004delving}. Mordern CAC approaches often include artifical intelligence (AI) since large language models (LLMs) like ChatGPT have demonstrated impressive capabilities in natural language processing tasks. However, these generative AI models tend to perform poorly when applied to clinical coding \cite{soroush2024large, falis2024can}, perhaps largely explained by the vast and intricate label space of ICD codes (with thousands of specific options) and the need for domain-specific expertise and lack of localised clinical data for training purposes. However, recent studies demonstrate that in some extremely limited instances characterized by narrowly defined label-space or clinical problem, ChatGPT can still perform well \cite{abdelgadir2024ai, wang2024validation}.

Even though complete or full automatic code assignment using current methods is nearly impossible, semi-automatic solutions present a practical and effective alternative. Instead of automatic assignment, these tools provide a list of code suggestions, while still relying on human oversight and control over the final coding decisions. However, to date, most studies evaluating CAC systems are conducted in controlled, lab-based environments, which limits their applicability to real-world clinical settings. While these studies provide valuable insights into the potential of CAC tools, they often fail to capture the complexities of real-life settings. This gap underscores the need for more field-based research with users. The current study was conducted with health staff using our CAC tool, Easy-ICD, that produces smart suggestions based on information in the Electronic Health Record (EHR).

\subsection{Related work}

Although research on CAC tools has a long history \cite{10.1145/243199.243276, carlo2010aligning}, much of the significant progress is only very recent, and can be attributed to the recent advances in deep learning methods \cite{dai2024evaluating,chen2021automatic}. These new methods are better adept at dealing with the challenges of the large label space (tens of thousands of codes), unbalanced datasets and long text sequences, even though they also introduce new challenges such as explainability \cite{dolk2022evaluation}. While deep learning is a key component in recent studies, ensembles combining multiple methods are quite common \cite{ZAMBETTA2024105462, chomutare_fuzzy, chen2022automatic}. A review by Yan et al. \cite{yan2022survey} provides an overview of how methodologies evolved from purely rule-based methods to neural network-based methods. It further demonstrates how renewed interest in the problem spans different training datasets in not only major languages such as English and Chinese, but also French, Italian, Hungarian, German and Spanish \cite{ZAMBETTA2024105462} \cite{mittermeierautomatic}. Increased research activity suggests that CAC systems have the potential to streamline the coding process and reduce the burden on clinical coders, especially when processing high volumes of patient records.

However, the evidence regarding the overall utility of CAC systems is not conclusive. Some studies have reported improved accuracy when a CAC system is used but without any time savings, and others have also reported time savings when a CAC system is used but without improving accuracy. For instance, Zhou et al. \cite{zhou2020construction} used regular expressions to implement a CAC system to enable automatic ICD-10 coding. The tool was developed and used during 16 months 2017-2018. 160,000 codes were automatically assigned and compared with manual coding. The finding was that the automatic system was 100 times faster in ICD-10 coding compared to manual coding, while still maintaining good coding quality. The $F_1$-score of the system was relatively low, around 0.6086.

In contrast, another study by Chen et al. \cite{chen2021automatic} constructed an automatic coding tool using the BERT (Bidirectional Encoder Representations from Transformers) algorithm trained on patient records from one Taiwanese hospital. The discharge summaries contained 14,602 labels in total. The ICD-10 auto-coding tool predicted ICD-10 codes with the best $F_1$-score of between 0.715 and 0.618. The tool was used in a user study where the tool did not really decrease coding time but the coding quality increased significantly from a median $F_1$-score of between 0.832 to 0.922.

While heterogeneity in the different studies partially explains the inconclusive evidence, there are some factors that are not fully explored in the literature, including complexity of the clinical texts. It is conceivable the complexity of text is a major factor when considering how a human versus a machine performs. In this study, we focus on text complexity, in an attempt to explore the conditions under which CAC systems are most effective. 

\subsection{Objective}
This study seeks to provide evidence on the potential of AI to enhance clinical coding practices, thereby improving operational efficiency and data quality in healthcare. We evaluate the effectiveness of the tool (see Fig~\ref{easy-icd-tool} for a screen dump of this tool), under two conditions: (i) when the clinical notes are complex versus (ii) when they are simple. We denote complexity by the number of tokens or words in each clinical note, rather than by complexity of the medical case. Word count is a simplified, yet also more objective measure that is effective, since longer texts generally require more time to read than shorter texts. We aim to answer the following research question: Does the Easy-ICD tool increase the speed and quality of ICD-10 coding compared to traditional manual coding for both complex and simple Swedish discharge summaries?.

\section{Methods}

We followed the CONSORT-AI template \cite{liu2020reporting} for reporting our study, which is a CONSORT extension for studies reporting AI-related interventions.

\subsection{Trial design}

The study is designed as a 2x2 crossover randomised controlled trial, using the test of equivalence as our framework, and an allocation ratio of 1:1. 

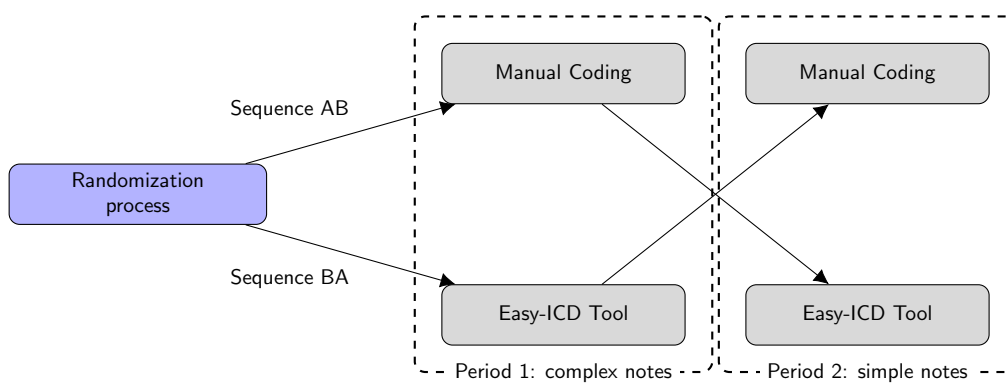
\begin{figure}[ht]
		\centering
		\begin{tikzpicture}[node distance=2cm,
			every node/.style={fill=white, font=\sffamily}, align=center, scale=0.8, transform shape]
			\node (start) [activityStarts,text width=4cm] {Randomization \\ process};
			\node (sequence1) [startstop, above of=start, xshift=7cm] {Manual Coding};
			\node (sequence2) [activityRuns, below of=start, xshift=7cm] {Easy-ICD Tool};
			\node (sequence3) [startstop, right of=sequence1, xshift=3cm] {Manual Coding};
			\node (sequence4) [activityRuns, right of=sequence2, xshift=3cm] {Easy-ICD Tool};
			
			\begin{scope}[on background layer]
			\node[draw, thick, dashed, rounded corners, inner xsep=1em, inner ysep=1em, fit=(sequence1) (sequence2)] (period1) {};
			\node[fill=white!20] at (period1.south) {Period 1: complex notes};
			\node[draw, thick, dashed, rounded corners, inner xsep=1em, inner ysep=1em, fit=(sequence3) (sequence4)] (period2) {};
			\node[fill=white!80] at (period2.south) {Period 2: simple notes};
 			\end{scope}
 			
			\draw [->] (start)  -- node [xshift=-1cm,yshift=.4cm]{Sequence AB}(sequence1) ;
			\draw [->] (start)  -- node [xshift=-1cm, yshift=-.4cm]{Sequence BA}  (sequence2);
			\draw [->] (sequence1) -- (sequence4);
			\draw [->] (sequence2) -- (sequence3);
		\end{tikzpicture}
		\caption{Crossover study design illustrating the AB$|$BA sequences, where Period 1 contains 65\% of the total word count and Period 2 contains 35\% of the total word count.}
		\label{fig:crossover_design}
	\end{figure}

The study design is illustrated in Fig~\ref{fig:crossover_design}, where participants alternate between using Manual Coding and the Easy-ICD tool to code complex and simple notes, with random assignment to different sequences to control for period and order effects. We denote complexity based on the number of words in the clinical note and not complexity of the medical case, where Period 1 contains 65\% of the total words and Period 2 contains only 35\%. This design helps assess the impact of the Easy-ICD tool on coding performance, in terms of accuracy and speed, by controlling for the complexity of the coding task and balancing the order of interventions.

 Participants were randomized into two groups (Sequence AB and Sequence BA) to ensure that each participant experiences both interventions in different orders. This ensures that each participant codes both complex and simple notes, but experiences the interventions in a different order to balance out any learning effects or biases related to note complexity. There were no important changes to methods after the trial commenced.

\subsection{Participants}
\subsubsection{Eligibility criteria for participants}

Participants were required to have prior experience in medical coding or relevant healthcare documentation to ensure familiarity with the coding tasks. The participants were nurses, coding experts, and physicians from Sweden and Norway. The focus was on professional coders, who usually are nurses trained in coding. Individuals with any medical condition or cognitive impairment that could affect their ability to complete the coding tasks were excluded from participation.

\subsubsection{Settings and location where the data was located}

The study was organised online, participants were recruited in Norway and Sweden. Each participant received a web link to the study and was allowed to complete it using a device of his/her choice whenever it was convenient. 

\subsection{Interventions}

\begin{figure}[ht]
\centering
\includegraphics[scale=0.54]{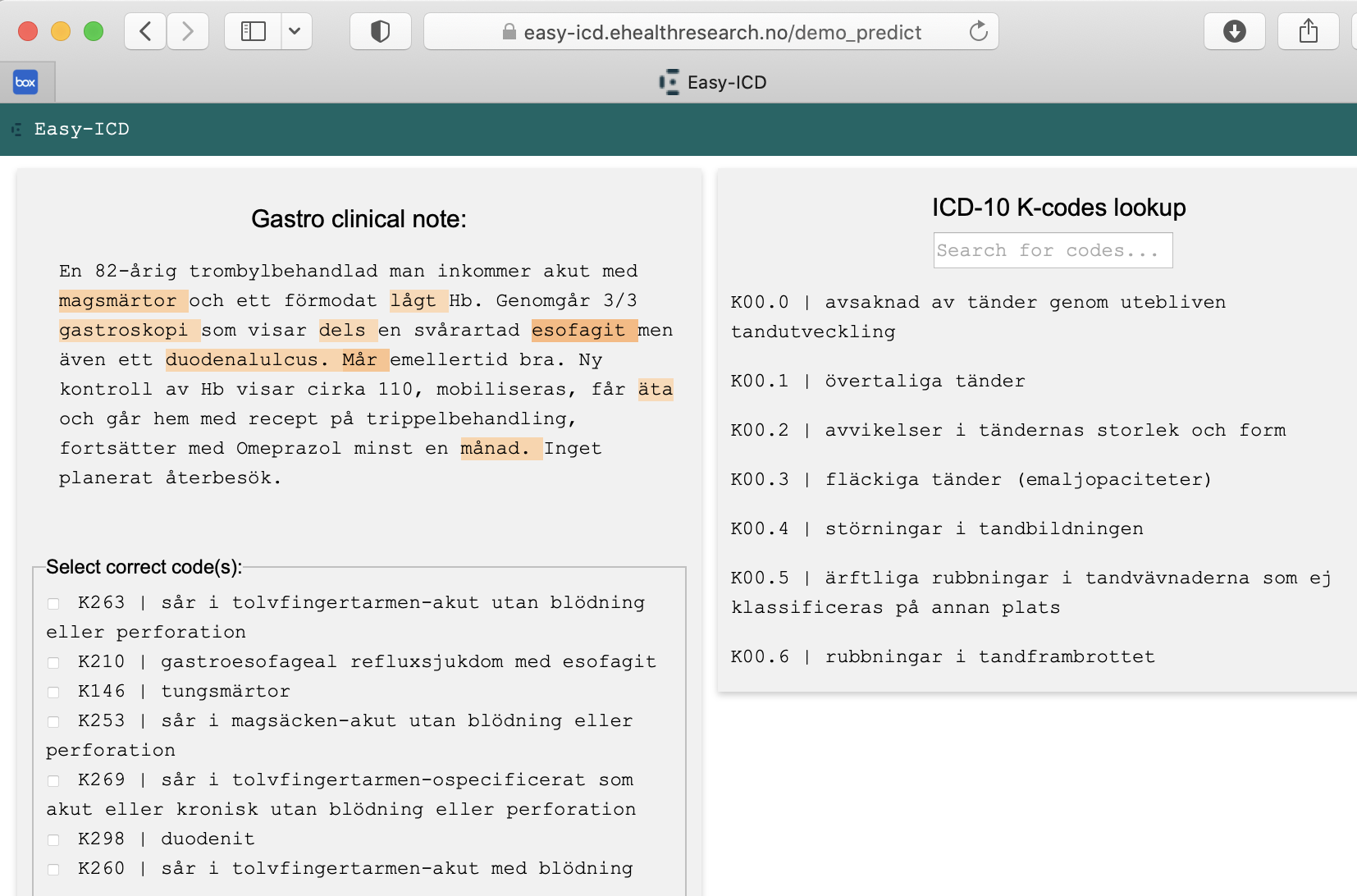}
\caption{Screen dump of the Easy-ICD DEMO.}
\label{easy-icd-tool}
\end{figure}

The coding tool, Easy-ICD, was designed as a web application using a typical Model View Controller (MVC) software design pattern \cite{garcia2023mvc}. The model has two components; a BERT-based clinical LLM as described in \cite{lambroudis2023using_icd_10_corpus}, and a fuzzy logic component described in \cite{chomutare_fuzzy}. The view is composed of custom HTML5 templates, while the controller that connects the user interface is based on Flask, a Python framework, see figure \ref{easy-icd-tool}, for a screen dump of the DEMO.

Data used for the training of the LLM model behind Easy-ICD-tool is the Stockholm EPR Gastro ICD Pseudo Corpus II\footnote{This research has been approved by the Regional Ethical Review Board in Stockholm Dnr 2007/1625-31/5 with the amendment by the Swedish Ethical Review Authority Dnr 2022-02386-02.
} fine tuned on the SweDeClin-BERT\footnote{This research has been approved by the Swedish Ethical Review Authority under permission no. 2019-05679. The use of SweDeClin-BERT with the amendment no 2022-02389-02.} Swedish clinical language model \cite{lambroudis2023using_icd_10_corpus}. A subset of 20 clinical notes from discharge summaries were re-coded by a senior Swedish coder for the purposes of quality-assuring the notes used in this study, that is, creating the gold standard. 

To use our solution, the participants needed a good understanding of the coding process. This usually means they have had some training in coding. For a given chunk of text, our solution suggests 5-7 codes, from which the participants have to select one or more codes. Participants may also choose not to select any code if they feel the suggestions do not contain any appropriate code. The number of codes the solution suggests (5-7 codes) was based on our discussion with clinical coders during the requirements gathering phases of the project. The solution's output was expected to help the coder identify correct codes more efficiently.

During this experimental study, participants were not allowed to be disrupted; we instructed them to take the task in one sitting. If they were disrupted, they were to quit the study immediately. To help the participants adhere to the requirements of the study, information redundancy was used; first in the consent form and also in a short video just before the coding task began.

\subsection{Outcomes}
There were two primary outcome measures in this study; the time used in assigning codes, and the accuracy of the coding. Time was logged as the participants navigated through the study by clicking the "Next" button. In terms of accuracy, we measured how the participants performed, compared to the gold standard.

One important change to the trial outcome was that we originally planned to measure usability of our solution as a secondary measure. We then considered that by the time participants reached the usability survey, they would likely be exhausted, which might have had an impact on the result. Since we had some usability feedback during the co-design process with clinical coders, we decided to conduct a separate usability study at a later date. However, we used star-rating of usefulness as an indirect qualitative gauge of user satisfaction with the system.

\subsection{Sample size}

To calculate the sample size for the two-sided tests used in our study, particularly for the Mann-Whitney U test (a non-parametric test), we needed to consider several factors, such as the desired power of the study (80\%), medium effect size (0.5), significance level (0.05). The significance level ($\alpha$) is the probability of a Type I error; rejecting the null hypothesis when it is true, while the power (1 - $\beta$) is the probability of correctly rejecting the null hypothesis when it is false (avoiding a Type II error). For the effect size, we used Cohen's d for Mann-Whitney U test, which quantifies the difference between groups. Given that smaller effect sizes require larger samples, we used  medium effect size. 

For a rough sample size estimate using Cohen’s effect size (d), the formula for the sample size per group (\( n \)) in a non-parametric context can be approximated using the following formula:

\[
n = \frac{Z_{\alpha/2} + Z_{\beta}}{d^2}
\]

Where:
\begin{itemize}
    \item  \( Z_{\alpha/2} \) is the Z-score for the desired significance level (for $\alpha$ = 0.05, Z = 1.96),
    \item \( Z_{\beta} \) is the Z-score for the desired power (for 80\%, Z = 0.84),
    \item \( d \) is the estimated effect size (Cohen's d).
\end{itemize}

Given that we assume a medium effect size \( d = 0.5 \), power of 80\%, and $\alpha$ of 0.05, approximately 32 observations per group would be needed in each of Period 1 and Period 2. Recruitment strategies included contacting hospitals and health authorities, announcing the study at coding seminars and conferences, and utilizing popular science media, social media, and other electronic platforms for advertising and outreach.

\subsection{Randomization}

The randomization sequence was generated using a alternate switching between 0 and 1 to ensure randomness and minimize allocation bias, and participants were assigned to Group 1 (sequence AB) and Group 2 (sequence BA) in a 1:1 ratio. The randomization process was implemented as follows: after obtaining informed consent, the participant was randomly assigned the next number in the 0-1 toggle sequence to reveal the participant's group assignment. The group assignment was then recorded in the study log. Participants were not informed of their group allocation, but they were made aware that they would be allocated to either group. Due to the nature of the intervention, there was no allocation concealment.

\subsection{Blinding}

Blinding was not utilized in this study because the intervention involved a clearly distinguishable user interface, making it impossible to mask from the participants and investigators. The nature of our solution required participant awareness for proper adherence. Additionally, the primary outcomes were objective measurements not subject to observer bias, further reducing the need for blinding. Thus, while blinding is a valuable technique to minimize bias, its application was neither feasible nor necessary for the integrity and validity of our particular study.

\subsection{Similarity of interventions}

Participants interacted with two different user interfaces that were designed to be visually and functionally similar to ensure a consistent user experience across both intervention and control groups. Both interfaces featured the same layout, design elements, and core functionalities to minimize any potential confounding variables related to usability or user satisfaction.

The primary distinction between the two interfaces was the inclusion of an artificial intelligence (AI) feature in the intervention interface. This AI feature provided real-time suggestions and recommendations to the users based on the clinical text they are reviewing. Conversely, the control interface did not include this AI suggestion feature; users reviewed the clinical texts and assigned codes without receiving any automated suggestions. 

Maintaining similar user interfaces ensured that any differences observed between the two groups could be attributed to the presence or absence of the AI feature, rather than other extraneous factors related to interface design or functionality.

\subsection{Statistical methods}

In this study, we opted not to perform commonly used statistical analyses in crossover studies, which typically account for the dependency between periods. The primary reason for this decision is because we needed to consider the impact of clinical note complexity, and we assigned complex notes to Period 1 and simple notes to Period 2. Therefore, we treated the two periods as independent since the levels of difficulty of the clinical notes varied significantly between the two periods. It would have been inappropriate to assume that performance in one period could be directly compared to performance in the other, as this could introduce bias related to task difficulty rather than the intervention itself.

Treating the periods independently enabled us to evaluate the effect of each intervention (Manual coding vs. Easy-ICD) under different task conditions (complex vs. simple notes) without conflating these factors. If we assume independence between periods, we avoid the complexities introduced by possible carry over effects and ensure a clearer, unbiased assessment of each intervention within each period. 

 We used the Mann-Whitney U test to analyze coding time for each period separately. The Mann-Whitney U test was appropriate because our data had a non-normal distribution. Also, the test is robust to outliers, which allowed us to compare the performance between the two groups without assuming normality. This approach provided a more straightforward interpretation of the results, given the distinct nature of the periods. Statistical analyses were performed using statistical software R and StatsDirect (version 4.0.4). A confidence interval (CI) if 95\% and p-value of less than 0.05 was considered statistically significant.

\subsection{Data collection and management}
All participant interactions with the user interfaces were logged to ensure comprehensive data collection. Every action performed by the participants, such as clicks and selections where logged. To track the duration of tasks and response times, the exact times at which each action was performed was also logged. This also includes all AI-generated suggestions and whether they were accepted, modified, or ignored by the participants. This logging was conducted in real-time and stored securely on the server, ensuring that no data was lost and that it was immediately available for analysis.

Data management was handled with strict adherence to privacy and security protocols to protect participant confidentiality and data integrity. All communication had end-to-end encryption using transport layer security (TLS) \cite{cremers2017comprehensive}, a commonly used standard for securing internet communications using web browsers. To keep the data anonymous, personal identifiers were not collected. No Internet protocol (IP) addresses were collected and cookies were only used to track progress in the experiment. This ensured that individual participants could not be directly identified from the dataset. 

Even though the raw data is anonymous, access to the data was restricted to authorized personnel only. A secure login system was used to manage access rights. Regular data integrity checks were performed to ensure that the data was complete, consistent, and free from corruption. Regular backups of the data were taken to prevent data loss. These backups were stored in separate, secure locations and could be restored in case of data corruption or accidental deletion.

\section{Results}

\subsection{Participant flow}
This study utilized a 1:1 crossover design involving 17 participants from an initial invitation pool of over a 100 coders. Each participant interacted with both the intervention and control interfaces in two separate periods, ensuring that all participants experienced both conditions. The order of exposure to the interfaces was randomized to control for potential order effects.

Participants were randomly assigned to one of two sequences; the first sequence (A$|$B) using the control interface, followed by the intervention interface, and the second sequence (B$|$A) first using the intervention interface, followed by the control interface. Participants in each sequence switched over between Period 1 and Period 2.

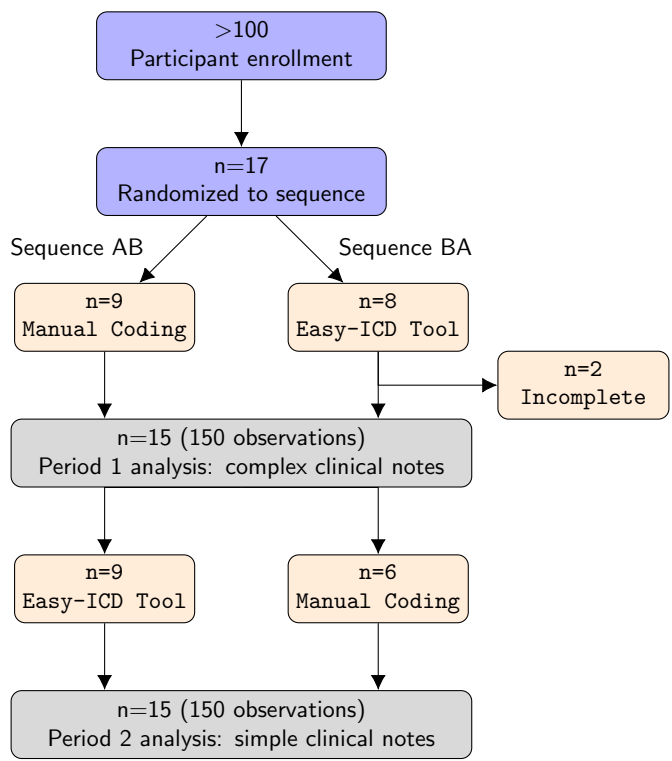
\begin{figure}[ht]
		\centering
		\begin{tikzpicture}[node distance=2cm,
			every node/.style={fill=white, font=\sffamily}, align=center, scale=0.9, transform shape]
			\node (enroll) [activityStarts, text width=4cm] {\textgreater 100 \\ Participant enrollment};
			\node (start) [activityStarts,text width=4cm, below of=enroll] {n=17 \\ Randomized to sequence};
			\node (sequence1) [process, below of=start,  xshift=-2cm] {n=9 \\ Manual Coding};
			\node (sequence2) [process, right of=sequence1, xshift=2cm] { n=8 \\ Easy-ICD Tool};
			\node (washout) [startstop, below of=sequence1,text width=6.5cm,  xshift=2cm] {n=15 (150 observations)\\ Period 1 analysis: complex  clinical notes};
			\node (sequence3) [process, below of=washout,  xshift=-2cm] {n=9 \\ Easy-ICD Tool};
			\node (sequence4) [process, right of=sequence3, xshift=2cm] {n=6 \\ Manual Coding};
			\node (end) [startstop, below of=sequence3,  text width=6.5cm,  xshift=2cm] {n=15 (150 observations) \\ Period 2 analysis: simple clinical notes};
			\node (dropped) [process, below of=sequence2, yshift=1cm, xshift=3cm] {n=2 \\ Incomplete};
			
			\draw [->] (enroll) -- (start);
			\draw [->] (start) -- node [xshift=-1.4cm]{Sequence AB} (sequence1) ;
			\draw [->] (start)  -- node [ xshift=1.4cm]{Sequence BA} (sequence2);
			\draw[->] (sequence2) |- (dropped);
			\draw [->] (sequence1.south) -- ([xshift=-2cm]washout.north);
			\draw [->] (sequence2.south) -- ([xshift=2cm]washout.north) ;
			\draw [->] (washout.south) -| (sequence3.north);
			\draw [->] (washout.south) -| (sequence4.north);
			\draw [->] (sequence3.south) -| ([xshift=-2cm]end.north);
			\draw [->] (sequence4.south) -| ([xshift=2cm]end.north);
		\end{tikzpicture}
		\caption{Crossover study design illustrating the AB$|$BA sequences, where Period 1 contains 10 complex clinical notes and Period 2 has an additional 10 simple clinical notes, for a total of 300 observations in this study.}
		\label{fig:crossover_flow}
	\end{figure}

As illustrated in the flow diagram in Fig~\ref{fig:crossover_flow}, of the 17 participants who enrolled, 15 completed both periods of the crossover study with valid data. The data from these 15 participants were included in the final analysis. Since each participants coded 20 clinical notes, there was a total of 300 observations or coded texts.

\subsection{Losses and exclusions}

 During the data collection, one participant's data was deemed invalid due to incomplete data logging, and the other participant only generated a random sequence but went no further. 
 
\subsection{Recruitment}
Participants of the user study where recruited through acquaintances from our professional network in Sweden and the Diagnosis-related Group (DRG) Forum, a network of professionals dedicated to enhancing the practice of DRG classification in Norway. The participants where given a link to the user study along with the instructions. They where also given a digital instruction of the study in a video tutorial right after the consent form.

Recruitment commenced November 2023 and continued on a rolling basis until May 2024. We had to stop the recruitment because the project funds had run out and it was difficult recruiting more participants. This is because our target audience are practising professional coders, and our speculation was that uncertainty regarding privacy may have resulted in the relatively low response rate. We also have some anecdotal evidence of coders expressing concerns that new AI systems may take over their jobs, and this may have caused some anxiety among potential participants.

\subsection{Baseline data}

Table~\ref{tab:user_char} shows the baseline demographic and characteristics of the participants. There were seven Norwegian and eight Swedish coders. The majority (n=11) had more than 5 years coding experience, while only one had less than one year's experience.

	\begin{table}[ht]
		\caption{Characteristics of participants}
		\centering
		\begin{tabular}{|l|l|l|}
			\hline
			 \textbf{Clincial coding experience (years)} & \textbf{Language}&\textbf{count}\\[1ex]
			\hline
			\textgreater5& Norwegian & 5\\ 
			\textgreater5&Swedish & 6\\ 
			1-5& Norwegian& 2\\ 
			1-5&Swedish & 1\\ 
			\textless1&Swedish&1 \\ 
			\hline
		\textbf{Total}&&\textbf{15}\\
			\hline
		\end{tabular}   
		\label{tab:user_char}
	\end{table}

\subsection{Numbers analysed}

The total number of participants were 15, with Group 1 consisted of 9 participants and Group 2 consisted of 6 participants (see ~Fig.\ref{fig:crossover_flow}).

\subsection{Outcomes and estimation}

\subsubsection{Median coding time}

\begin{figure}[ht]
\centering
\includegraphics[scale=.2]{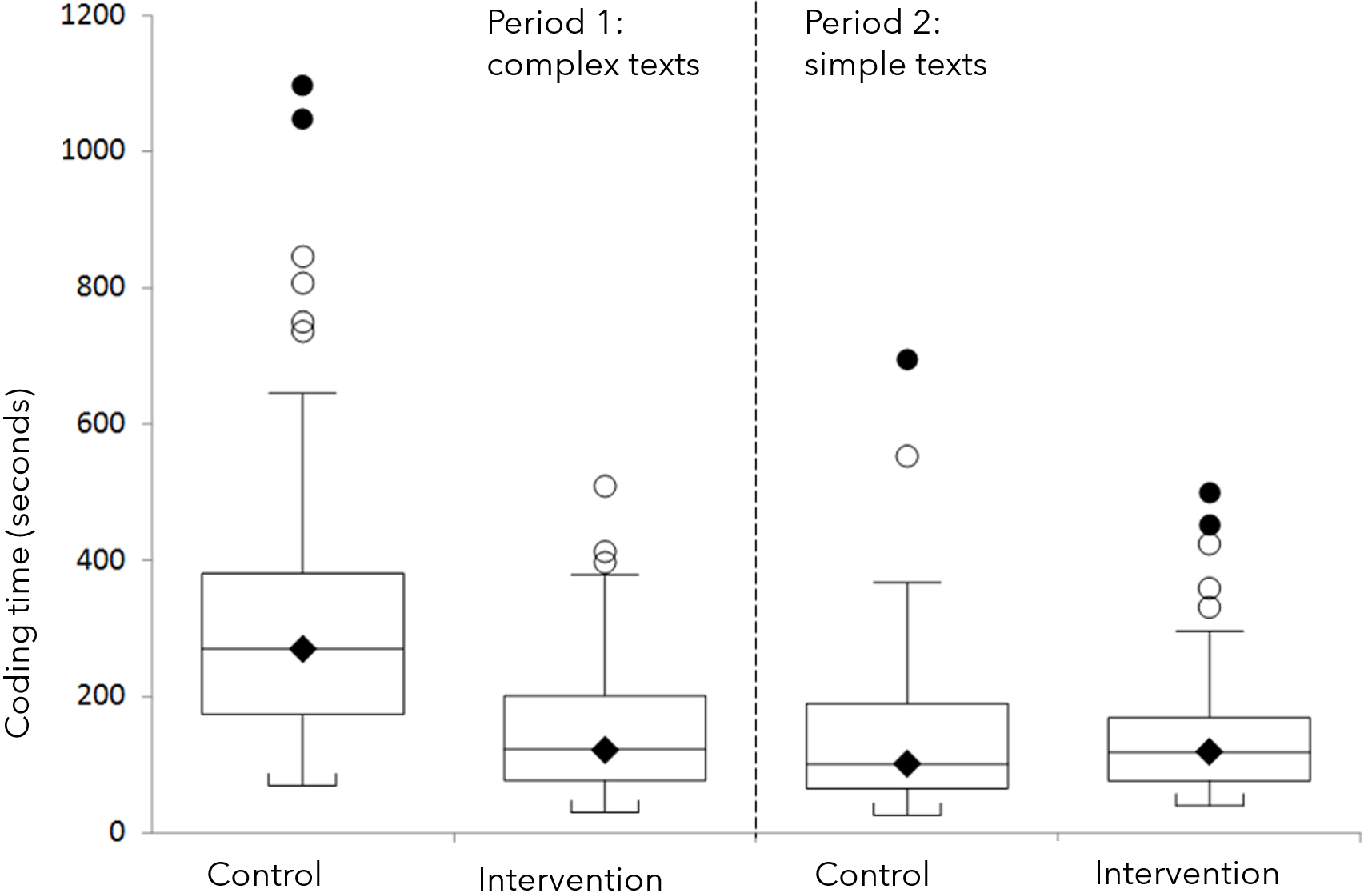}
\caption{Clinical coding time (sec.) for complex and simple clinical texts.}
\label{fig:boxplot}
\end{figure}

 In terms of the descriptive statistics, the control had a mean coding time of 242.6 sec. (192.3 SD) while the intervention had 147 sec. (94.5 SD). Due to the presence of outliers and the non-parametric nature of the data distribution, we used the Mann-Whitney U test to analyze the data. The Mann-Whitney U test is a non-parametric test that does not require the assumption of normality and is less sensitive to outliers and imbalanced sample sizes. Data distribution and outliers are illustrated in the box plot in Fig~\ref{fig:boxplot}. The test is specifically designed to assesses whether the median difference between the observations is significantly different from zero. This allowed us to effectively compare the performance metrics between the two interventions for each participant. 

\begin{table}[ht]
	\centering
	\caption{Mann-Whitney U test for code assignment time duration}
	\label{tab:mann_whitney}
	{
		\begin{tabular}{llrrrrrr}
			\toprule
			\multicolumn{1}{c}{} & \multicolumn{1}{c}{} & \multicolumn{1}{c}{} & \multicolumn{1}{c}{} & \multicolumn{2}{c}{95\% CI} \\
			\cline{5-6}
			 & Sample size & 2-sided $p$ & Median (diff.)  & Lower & Upper  \\
			\cmidrule[0.4pt]{1-6}
			Period 1: control & n=90 &  & $270$ &  &   \\
			Period 1: intervention & n=60 &  & $121.5$ & &   \\
                Complex notes (Period 1) & n=150 (total) & $p\textless0.0001$ & $(123)$ & $81$ & $164$  \\[1ex]
                \hline 
                Period 2: control & n=60 &  & $101.5$ &  &   \\
			Period 2: intervention & n=90 &  & $119$ &  &   \\
                Simple notes (Period 2) & n=150 (total) & $p=0.2455$ & $(-11)$ & $-34$ & $8$  \\
			\bottomrule
		\end{tabular}
	}
\end{table}

As shown in Table~\ref{tab:mann_whitney}, for the complex (longer) clinical texts in Period 1 (1-10), the median difference in coding time duration between using our solution and not using our solution was 123 seconds at 95\% CI(81:164) and a two-sided p\textless.0001. This represents an approximate 46\% reduction in median coding time when our solution was used. For the shorter clinical texts in Period 2 (11-20), the median difference was 11 seconds at 95\% CI (-34:8) and a two sided p\textless.2455, indicating an insignificant difference. 

\subsubsection{Clinical coding accuracy}

Table~\ref{tab:f1} shows the weighted average metrics for coding performance, where we see a general increase in accuracy when our tool was used compared to when it was not. However, testing these outcomes for significance based on two independent proportions, we discovered that the improvement in performance was not statistically significant, as shown in Table~\ref{tab:proportions}.

	\begin{table}[ht]
		\centering
		\caption{Weighted average metrics for coding performance; with and without Easy-ICD.}
		\label{tab:f1}
		{
			\begin{tabular}{lrrrrr}
				\toprule
				 & Precision & Recall & f1-score & Accuracy & N  \\
				\cmidrule[0.4pt]{1-6}
				\emph{Complex notes} &&&&&\\
				Control  & $0.90$ & $0.62$ & $0.71$ & $0.62$ & $90$  \\
				Intervention & $0.90$ & $0.67$ & $0.74$ & $0.67$ & $60$  \\
				\hline
				\emph{Simple notes} &&&&\\
				Control  & $0.90$ & $0.60$ & $0.70$ & $0.60$ & $60$  \\
				Intervention & $1.00$ & $0.70$ & $0.79$ & $0.70$ & $90$  \\
				\bottomrule
			\end{tabular}
		}
	\end{table}

	\begin{table}[ht]
		\centering
		\caption{Two independent proportions for coding accuracy}
		\label{tab:proportions}
		{
			\begin{tabular}{llrrrrrr}
				\toprule
				\multicolumn{1}{c}{} & \multicolumn{1}{c}{} & \multicolumn{1}{c}{} & \multicolumn{1}{c}{} & \multicolumn{2}{c}{95\% CI} \\
				\cline{5-6}
				& Sample size & 2-sided $p$ & Accuracy(diff.)  & Lower & Upper  \\
				\cmidrule[0.4pt]{1-6}
				Period 1: control & n=90 & $$ & $0.62$ & $$ & $$  \\
				Period 1: intervention & n=60 & $$ & $0.67$ & $$ & $$  \\
				Complex notes (Period 1) & n=150 (total) & $p\textless0.4986$ & $(-0.044)$ & $-0.195$ & $0.114$  \\[1ex]
				\hline 
				Period 2: control & n=60 & $$ & $0.6$ & $$ & $$  \\
				Period 2: intervention & n=90 & $$ & $0.7$ & $$ & $$  \\
				Simple notes (Period 2) & n=150 (total) & $p=0.1692$ & $(-0.1)$ & $-0.255$ & $0.054$  \\
				\bottomrule
			\end{tabular}
		}
	\end{table}

\subsubsection{User satisfaction}
\begin{figure}[ht!]
\centering
\includegraphics[scale=.2]{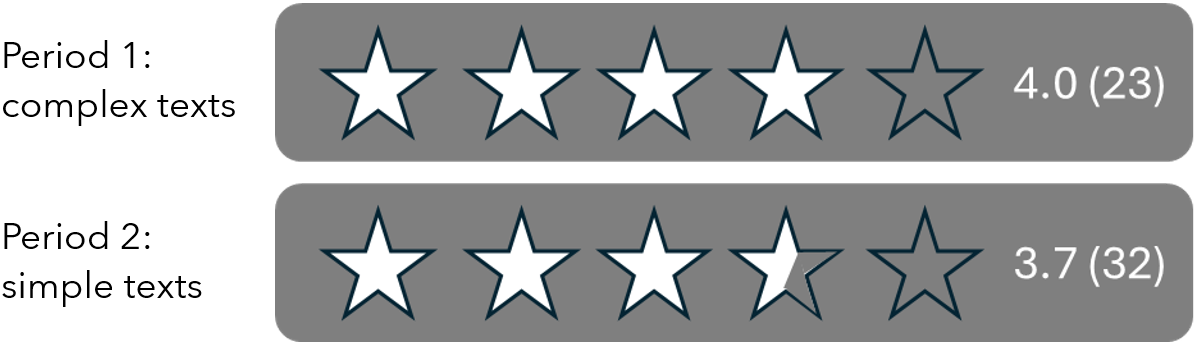}
\caption{Qualitative feedback in the form of 5-star rating for the suggested codes for complex notes (top) and simple notes (bottom) for the 55 submitted ratings out of a possible total of 150 notes that could have been rated.}
\label{fig:ratings}
\end{figure}

In terms of the user satisfaction with the provided suggestions, participants were slightly more satisfied with the suggestion during the complex notes compared to suggestions for simpler notes, as illustrated in star rating in Fig.~\ref{fig:ratings}. However, it should be noted that it is only about a third (n=55/150, 36.7\%) of the clinical notes whose suggestions were rated.

\subsection{Harms}
No harms were recorded.

\section{Discussion}

Results emerging from our study show that the Easy-ICD tool was most useful for complex or longer clinical notes, compared to simple or short clinical notes. These results are not surprising, since we initially hypothesised that such an AI assistive tool could be most useful for complex clinical notes, and that the varying reports in the literature could possibly be partially explained by this factor.

Our study confirms that tools such as our Easy-ICD have the potential to contribute to decision making and to reducing the excessive documentation burden on healthcare staff. Through automating the discharge summary coding process, such tools are expected to improve the speed and quality of the coding, thus facilitating more efficient healthcare delivery. Use of AI can result in more accurate ICD-10 codes, reducing the probability of errors and missing codes in clinical documentation. This automation not only saves time, but it also ensures consistency in coding practices; leading to improved patient care and streamlined administrative processes.

\subsection{Limitations}

 One limitation of the trial could be selection bias, as participants probably had much higher confidence and affinity to testing new technology, and they were recruited from specific pool of healthcare settings in Norway and Sweden. There is also a risk of order effect or learning effect bias in a crossover design where the sequence in which participants receive the intervention may influence their response or performance. For example, participants may become more proficient and improve their coding skills due to repeated exposure to the task or the introduction of the intervention. This could lead to better outcomes in the later period of the trial.
 
 Study findings may have been influenced by factors such as coder experience; a sample size meant we could not analyse the influence of experience on the results, and this could be an opportunity for further study.

 Another limitation is that the Easy-ICD tool only predict chapter XI codes, the so called K-codes. This is a very limited scope of the overall ICD-10 coding system. Perhaps including other chapters would have presented more realistic, and possibly also more challenging, scenarios for coders. We leave this question as a possible object of future inquiry.

\subsection{Generalizability}

The successful application of Easy-ICD to improve the efficiency of clinical coding in the study suggests that similar AI tools could be beneficial in healthcare systems in general. The findings of this study highlight the potential of AI interventions to improve clinical coding practices, regardless of geographical location or healthcare setting.

Additionally, the positive outcomes observed in this trial indicate that the benefits of AI in clinical coding are transferable and applicable to various healthcare settings. By demonstrating the effectiveness of Easy-ICD in optimizing coding processes and improving data quality, this study provides valuable information on the generalizability of AI interventions in healthcare.

This may also present opportunities for porting our tools to support ICD-11. Even though world governments are gearing towards implementing this new version, healthcare staff are less accepting of this new version. Assistive tools, such as the one presented in this study, may be the key to adoption and final transition to ICD-11.

However, we do acknowledge that the limitations we discussed have to be addressed, and more studies with bigger sample sizes are required before we can fully understand generalizability in this context.

\subsection{Interpretation}
The reduction in median coding time was only statistically significant for complex notes, suggesting that such assistive tools are comparatively more appropriate for complex tasks. We also noted that accuracy improved in both complex and simple notes, albeit not significantly, when our tool was used. This suggests that even though participants were quick to pick out codes for simple texts, those that used our tool had better accuracy even for simple texts, and this is something that needs further investigation with a larger sample size. 

However, it should also be noted that two possible explanations for the non-significant results on accuracy are that the code space was comparatively smaller (K-codes) and that the participants were highly motivated individuals with a high level of skill, and some of them were in leadership positions responsible for coding.

\section{Conclusion}
Our results highlight the potential of assistive tools, especially for more complex clinical coding tasks. This has important practical implications for the use of AI in clinical coding as these findings demonstrate that assistive technology can be effective productivity tools that reduce the excessive burden of administrative documentation. This is particularly relevant in healthcare where manpower is limited and accurate task completion is critical. Overall, the study demonstrates the value of AI in augmenting human performance, providing a compelling case for the broader adoption of AI-assisted interfaces to enhance productivity and accuracy in clinical coding. Globally, such AI tools have the potential to ease the adoption of more complex and detailed classification systems like ICD-11.

\section{Other information}

\subsection{Registration} 
https://clinicaltrials.gov/study/NCT0628686

\subsection{Protocol} JMIR Protocols \cite{chomutare_2024_user_study_protocol}
\subsection{Funding}

This research is funded by the ClinCode project through the Research Council of Norway, grant no. 318098., awarded to HD. The funder had no role in study design, data collection and analysis, decision to publish, or preparation of the manuscript.

\subsection{Data availability}
The datasets generated during the current study are available in the GitHub repository at \href{https://github.com/icd-coding/clincode_demo}{https://github.com/icd-coding/clincode\_demo}. The 20 clinical texts used in this study are not publicly available, but may be made available to qualified researchers on reasonable request from the corresponding author.

\subsection{Code availability}
A demonstration of Easy-ICD is located at \href{https://easy-icd.ehealthresearch.no}{https://easy-icd.ehealthresearch.no}.

\bibliographystyle{unsrtnat}
\bibliography{references}
\end{document}